\documentclass[preprint,preprintnumbers,amsmath,amssymb]{revtex4}
\usepackage{amsfonts}
\usepackage{amssymb}
\usepackage{latexsym}
\usepackage{graphicx}
\usepackage{psfig}

\begin{document}
\draft


\title{Mutual information and electron correlation in momentum space}
\author{Robin P. Sagar  and Nicolais L. Guevara \footnote{Present address: \it
 Instituto de Ciencias Nucleares, Universidad Nacional Aut\'onoma de M\'exico, D.F., 04510 M\'exico. }
   } 
\address{Departamento  de  Qu\'{\i}mica,  Universidad  Aut\'onoma
Metropolitana Apartado Postal 55-534, Iztapalapa, 09340 M\'exico D.F., M\'exico}
\date{\today}


\begin{abstract}
Mutual information and information entropies in momentum space are proposed as measures of the non-local aspects of information. Singlet and triplet state members of the helium isoelectronic series are employed to examine Coulomb and Fermi correlation, and their manifestations, in both the position and momentum space mutual information measures. The triplet state measures exemplify that the magnitude of the spatial correlations relative to the momentum correlations, depends on, and may be controlled by the strength of the electronic correlation. Examination of one and two-electron Shannon entropies in the triplet state series yields a crossover point, which is characterized by a localized momentum density. 
The mutual information density in momentum space illustrates that this localization is accompanied by strong correlation at small values of $p$.
\end{abstract}
\maketitle
\newpage


\section{Introduction }

The electron correlation effect lies at the center of many interesting phenomena that electronic systems display. It has broad ramifications, which range from the atoms and molecules of traditional quantum chemistry, and transcends to the truly many body systems of condensed matter. At the present time, electron correlation is thought to be responsible for phenomena such as high temperature superconductivity, and quantum phase transitions, due to quantum fluctuations arising from the uncertainty principle. The correlation among atoms in a Fermi gas is responsible for fermionic condensates.

The term electron correlation is usually reserved for the mutual Coulombic repulsion between electrons, however there is also the Fermi or exchange correlation between electrons of like spin, which is a consequence of the antisymmetry inherent in the wave function of a system of indistinguishable electrons. A detailed understanding of these effects, and the manner in which electrons avoid each other, is necessary if one hopes to understand the phenomena associated with them. For example, Fermi correlation has been related to the existence of shell structure in atoms and chemical binding or pairing in molecules \cite{baderjacs}. In fact, it has been shown to be responsible for the stability of matter, and the physical properties of our universe as we know it \cite{lieb}.

To understand the physical effects of electron correlation, it would first be helpful if one had a quantitative measure which could be used to determine if a particular system is strongly or weakly correlated. The correlation energy, defined as the difference between the exact non-relativistic and Hartree-Fock energies, is normally used in quantum chemistry as a measure of electron correlation. There are, however, other measures that have been introduced, such as the correlation coefficient \cite{kutz}, the degree of correlation \cite{grobe}, the Jaynes or correlation entropy \cite{esquivel,ziesche}, the Shannon entropy sum \cite{guevarapra}, and mutual information \cite{sagar}, among others.

The entropic and information based measures mentioned above have their foundations in information theory \cite{shannon}. In recent years, information theory has been utilized to study the electron correlation problem \cite{esquivel,guevarapra,hojpb,ramirez,ziesche1,ziesche2,guevarajcp2,romerajcp}. The relationship between electron correlation and entanglement \cite{huang} or non-separability of quantum systems, as studied by $quantum$ information theory, has also been discussed in the literature. The move to quantum information theory, based on density matrices, has been made because classical information theory, based on probability distributions, ignores the non-local aspects of information due to the superposition principle. This is related to the so-called phase relationships or wave interference effects, necessary if one wishes to examine the fully quantum behavior of information. 
We will argue later on that the non-local aspects of information, or correlation, from a position space perspective, can be recovered by examining mutual information in momentum space. Such an examination also serves practical purposes since correlation in momentum space is responsible for the small kinetic energy component of the exchange-correlation energy in Kohn-Sham density functional theory. There is also the promise that electron correlation in momentum space can be studied experimentally \cite{boeyen}.

The one-electron Shannon measure of uncertainty, or information entropy, has been studied for atomic systems in both the position and momentum representations \cite{gadre,hojpb}. They are defined as 
\begin{equation}
S_\rho= - \int \rho({\bf r}) \ln \rho({\bf r}) d {\bf r}
\label{shanr}
\end{equation}
\begin{equation}
S_\pi= - \int \pi({\bf p}) \ln \pi({\bf p}) d {\bf p},
\label{shanp}
\end{equation}
where $\rho(\bf r)$ and $\pi(\bf p)$ are the one-electron charge and momentum density respectively, and both densities
$\pi(\bf p)$ and $\rho(\bf r)$ are normalized to N, the number of electrons in the system. As measures of uncertainty or localization, the entropies are connected by an entropic uncertainty relationship \cite{gadre,guevarajcp1,bbm}
\begin{equation}
 0 \leq 3(1+ \ln \pi) \leq S_{\rho}^{u}+ S_{\pi}^{u}=S_{t}^{u},  
\label{guevara03}
\end{equation}
where the superscript, $u$, denotes unity normalized densities. Smaller values of the entropies are associated with localization of the underlying densities, and larger values with delocalization.

Likewise, two-electron Shannon entropies \cite{guevarajcp1,amoxvilli} are defined as
\begin{equation}
S_{\Gamma}= - \int \Gamma({\bf{r}}_1, {\bf{r}}_2) \ln \Gamma({\bf{r}}_1, {\bf{r}}_2) d{\bf{r}}_1 d{\bf{r}}_2
\label{shanr1}
\end{equation}
\begin{equation}
S_{\Pi}= - \int \Pi({\bf{p}}_1, {\bf{p}}_2) \ln \Pi({\bf{p}}_1, {\bf{p}}_2) d{\bf{p}}_1 d{\bf{p}}_2,
\label{shanp1}
\end{equation}
where $\Gamma({\bf{r}}_1, {\bf{r}}_2)$ and
$\Pi({\bf{p}}_1, {\bf{p}}_2)$
are the spinless two-electron densities, in position and
momentum space respectively, normalized to $N(N-1)$. These two-electron entropies also satisfy an entropic uncertainty relationship \cite{guevarajcp1},
\begin{equation}
 0 \leq 6(1+ \ln \pi) \leq S_{\Gamma}^{u}+ S_{\Pi}^{u}=S_{T}^{u}  
\label{bial1001}
\end{equation}
which is valid for an even number of electrons. The relationship between entanglement and entropic uncertainty relationships has recently been commented on in the literature \cite{guhne}.

The mutual information in position space,
\begin{equation}
I_r=\int \Gamma^u({\bf{r}}_1, {\bf{r}}_2) \ln {\Huge [}{\Gamma^u({\bf{r}}_1, {\bf{r}}_2) \over \rho^u({\bf{r}}_1)\rho^u({\bf{r}}_2)}{\Huge ]} d{\bf{r}}_1 d{\bf{r}}_2 = 2S^u_{\rho}-S^u_{\Gamma} \geq 0,
\label{mir}
\end{equation}
has been studied as an electron correlation measure \cite{sagar}, and along with its momentum space counterpart,
\begin{equation}
I_p=\int \Pi^u({\bf{p}}_1, {\bf{p}}_2) \ln {\Huge [}{\Pi^u({\bf{p}}_1, {\bf{p}}_2) \over \pi^u({\bf{p}}_1)\pi^u({\bf{p}}_2)}{\Huge ]} d{\bf{p}}_1 d{\bf{p}}_2 = 2S^u_{\pi}-S^u_{\Pi}
 \geq 0,
\label{mip}
\end{equation}
satisfies an uncertainty-type relationship \cite{sagar}
\begin{equation}
I_r + I_p \geq 0.
\label{unceri}
\end{equation}
Mutual information measures the interdependence or correlation between two variables using a Hartree-type state as its reference. 
This correlation between pairs of variables may also be regarded as the extent of pairing, or the collective pair behavior in the system. In the limit of strong correlation, the two particles would behave in unison, as a pair. Thus, mutual information may also be interpreted as the extent of pairing in a system.

At this point, it would be worthwhile to comment on the physical and mathematical nature of the entropic uncertainty relationships. These inequalities are a consequence of the Heisenberg uncertainty principle, i.e. smaller uncertainty (localization) in one space represents larger uncertainty (delocalization) in the other space due to the bounds in Eqs. (\ref{guevara03}) and (\ref{bial1001}). 

Physically, the usual interpretation (in one-electron atomic systems) is that localization of the electron's position results in an increase in kinetic energy and a delocalization of the momentum density. One can think of the nuclear potential as controlling the extent of this localization/delocalization. A weak nuclear potential would imply a delocalized electron charge density and hence a localized electron momentum density. The uncertainty principle in one-electron systems may be interpreted as a statement about the correlation between measurements of the co-ordinates ${\bf r}$ and momenta ${\bf p}$ (interaction with measuring device). A natural question is then, What is the connection between electron correlation in multi-electron systems and the entropic uncertainty relationships, particularly at the two-electron level, given by Eq. (\ref{bial1001}) ?

It is also important to note that Eq. (\ref{guevara03}) is a formulation of an entropic uncertainty relationship in terms of the Shannon entropies of the one-electron densities of a N-electron system. These one-electron densities correspond to antisymmetric wave functions which would also include Coulomb correlation depending on the level of approximation used. Thus, Eq. (\ref{guevara03}), for the entropy sum, is also a statement about the correlation among electrons, which is present in systems with more than one electron \cite{guevarapra}. The entropy sum of confined atoms has recently been discussed \cite{sen}.

A conceptually simple model to explain the role of electron correlation (Fermi and Coulomb) in the uncertainty relations is that correlation induces a delocalization in the position space charge densities since electrons avoid each other. The delocalization here is $relative$ to another lesser correlated charge density. The result of this delocalization is a corresponding localization in the momentum space density, due to the uncertainty principle. 
Thus $S_{\rho,\Gamma}$ should increase with correlation while $S_{\pi,\Pi}$ decrease. Such a model is also consistent with that of the one-electron systems if one accepts that the effect of the electronic interaction can be translated into an effective nuclear charge (shielding).

Another aspect of the entropic uncertainty relations is that these provide a lower bound on the sum of the entropies, but no explicit condition
on the relationship between the individual entropies, although lower bounds for the individual entropies 
have been formulated \cite{gadre2,chatz}. For example, What would one expect for the relative magnitudes of $S_{\rho}$ compared to $S_{\pi}$, and $S_{\Gamma}$ with $S_{\Pi}$? And what about the relative magnitudes of $I_r$ and $I_p$ in Eq.(\ref{unceri})? Would these change as physical aspects of the system, such as electron correlation or the nuclear potential, are varied?

We wish to advise the reader that we have used a loose interpretation of the uncertainty principle in the above. Delocalization in the position space does not necessarily imply localization in momentum space, from the standpoint of the  entropic uncertainty relationships, since these provide a lower bound, not an upper bound. To be more precise, one should use the concept of complementarity between position and momentum, and its connection to the uncertainty principle. For our purposes, we shall use the uncertainty principle as being equivalent to a statement about complementarity.

Mathematically, the localization-delocalization effect of the uncertainty principle is expressed by the Dirac-Fourier transform which connects the wave function in one space to the wave function in the other. The N-electron wave function in position space, $\Psi$, is related to its counterpart in momentum space, $\Phi$, by ,
\begin{equation}
\Phi ({\bf p}_1, \cdots, {\bf p}_N) = (2\pi)^{-{3N \over 2}}\int \Psi ({\bf r}_1, \cdots,{\bf r}_N)\exp(-i \sum _{j=1}^N {\bf p}_j.{\bf r}_j) d{\bf r}_1 \cdots d{\bf r}_N.
\label{ftrp}
\end{equation}
One can see from this relationship that each point in $p$-space contains contributions from $all$ points in $r$-space (and vice-versa). One may say that each point in $p$-space is $enfolded$ with all points in $r$-space (and vice-versa). 

The density matrix formulation of quantum mechanics, as used in electronic structure theory \cite{davidson}, is convenient since one may reduce from the N-electron problem by integrating over $(N-p)$ electronic variables, to yield the $p^{th}$ order reduced density matrix. This formulation allows one to retain the non-local aspects inherent in the wave function. The second and first order reduced density matrices are defined respectively as
\begin{equation}
\Gamma ({\bf r}_1,{\bf r}_2,{\bf r}_1^{\prime},{\bf r}_2^{\prime}) = N(N-1) \int 
\Psi ({\bf r}_1, {\bf r}_2, \cdots,{\bf r}_N)
\Psi^* ({\bf r}_1^{\prime}, {\bf r}_2^{\prime}, \cdots,{\bf r}_N)
d{\bf r}_3 \cdots d{\bf r}_N
\label{ODM2}
\end{equation}
and
\begin{equation}
\gamma ({\bf r_1},{\bf r_1}^{\prime}) = N \int
\Psi ({\bf r}_1, \cdots,{\bf r}_N)
\Psi^* ({\bf r}_1^{\prime}, \cdots,{\bf r}_N)
d{\bf r}_2 \cdots d{\bf r}_N.
\label{ODM1}
\end{equation}
%
We now examine the last equation. From this point on we suppress the index on one-electron quantities. When ${\bf r}={\bf r}^{\prime}$, one recovers the charge density, $\rho({\bf r})$, or diagonal element of the first order reduced density matrix, which provides the local behavior in $r$-space. The non-local behavior in $r$-space, (${\bf r}\neq {\bf r}^{\prime}$), is contained in the off-diagonal elements, and provides information about the quantum or wave-like nature. 

The first order reduced density matrix in momentum space is also connected to its position space counterpart by a Dirac-Fourier transform,
\begin{equation}
\gamma ({\bf p},{\bf p}^{\prime}) = (2\pi)^{-3} \int \gamma ({\bf r},{\bf r}^{\prime})
\exp[-i({\bf p} \cdot {\bf r} - {\bf p}^{\prime} \cdot {\bf r}^{\prime})] d{\bf r} d{\bf r}^{\prime}.
\label{ODMFT}
\end{equation}
Since ${\bf p}$ and ${\bf p}^{\prime}$ (from the integrations) contain contributions from all points in ${\bf r}$ and ${\bf r}^{\prime}$, i.e. both ${\bf r} = {\bf r}^{\prime}$ $and$ ${\bf r} \neq {\bf r}^{\prime}$, then $\gamma ({\bf p},{\bf p}^{\prime})$ and its diagonal element, $\pi ({\bf p})$, the momentum space density, must contain non-local behavior (${\bf r} \neq {\bf r}^{\prime}$). Thus momentum space measures, such as $S_{\pi}$, must contain the non-local behavior {\it in the r-space} representation. The same is also true for the position space measure, $S_{\rho}$. That is, it contains the non-local behavior in the $p$-space representation.  Such arguments are also valid for the second order reduced density matrix. Similar ideas to these have been previously presented \cite{ramirez1}. The relationship between local and non-local information in quantum information theory has also been a topic of discussion in the literature \cite{oppenheim}. 

The localization/delocalization phenomena contained in the uncertainty principle may also be interpreted from the standpoint of density matrices. With delocalization, or wave-like behavior, in a particular representation, one would expect significant contributions from the off-diagonal elements of the density matrices. This should translate into a more localized density in the reciprocal space.

$\rho({\bf r})$ and $\pi ({\bf p})$ are not strictly independent and do contain information about each other \cite{harriman} since there is a correlation between the two due to the uncertainty principle. Relationships between the two have been explored \cite{harbola}. Hence in associating $S_{\rho}$, $S_{\Gamma}$ and $I_r$ with local information in  $r$-space, and $S_{\pi}$, $S_{\Pi}$ and $I_p$ with non-local information, one should be aware that there is some mixing involved. However, the benefit gained from such a definition, i.e. position space corresponds to local information while momentum space corresponds to non-local information (or vice-versa), would be the connection to the entropic(information) uncertainty relationships in Eqs. (\ref{guevara03}), (\ref{bial1001}) and (\ref{unceri}). Thus these relationships could be interpreted as a connection between local and non-local information.

\subsection{The helium isoelectronic series}

In this study, we shall focus on correlation effects in the helium isoelectronic series. The nuclear charge, Z, is used as the parameter to control electron correlation. Electron correlation effects are largest for small Z, while they decrease with increasing Z, as the electron-nuclear interaction begins to dominate the electron-electron repulsion \cite{ziesche1}. Simple wave functions of the form
\begin{equation}
\Psi_S(r_1,r_2) = C_N(e^{-Z_1r_1}e^{-Z_2r_2} + e^{-Z_2r_1}e^{-Z_1r_2}),
\label{hewfns}
\end{equation}
can be used to represent the singlet ground states of the He series. $C_N$ is the normalization constant and $Z_1$ and $Z_2$ are variational parameters. These functions have been used to study the behavior of $I_r$ \cite{sagar}. Due to the symmetric spatial part of these functions, there is no Fermi correlation, and all the correlation is strictly Coulombic, through the presence of the coulomb hole. 

On the other hand, both Fermi and Coulomb holes are present in the excited triplet (1s2s) state, for which a wave function may be written as
\begin{equation}
\Psi_T(r_1,r_2) = C_N^{\prime}[e^{-Z_1^{\prime}r_1}e^{-Z_2^{\prime}r_2}(1-Z_2^{\prime}r_2) - e^{-Z_2^{\prime}r_1}e^{-Z_1^{\prime}r_2}(1-Z_2^{\prime}r_1)].
\label{hewfnst}
\end{equation}
Comparing the singlet and triplet entropic and information measures in the $r$-space and in $p$-space, one may comment on the differences or similarities in behaviors, due to purely Coulombic, or to Coulomb and Fermi correlation. A critical point, where the one-electron position space Shannon entropy becomes discontinuous, has been reported in the He isoelectronic series \cite{shi}. The physical effect of electron correlation in these systems, in position and momentum space, has been discussed in the literature \cite{thakkar,banyard,krause,youngman}. The differences between Fermi and Coulomb correlation in the helium triplet has been analyzed \cite{boyd}, and angular aspects of correlation  in the singlet and triplet states of helium have been compared \cite{ugalde}. 

The purpose of this paper is to study the mutual information and entropic measures, in position and momentum space, and to study their behavior with respect to Coulomb and Fermi correlation. We ask the question of how do Fermi and Coulomb correlation manifest themselves in the information and entropic measures, and what are the differences between Fermi and Coulomb correlation. Are there situations where electron correlation in position space (spatial correlation) is dominant over that in momentum space, ($I_r$ $>$ $I_p$), and vice-versa, and how does this depend on the strength of the correlation? Can this relative ordering be adjusted by varying the strength of the correlation? What does this tell us about the pairing mechanisms in each space? Atomic units are used throughout the paper.

\section{Results and Discussion}

\subsection{Mutual information}

All measures were calculated using spherically averaged densities and numerical integration. Momentum space densities were calculated by Dirac-Fourier transforming the wave functions in Eqs. (\ref{hewfns}) and (\ref{hewfnst}) into momentum space.
We present in Fig. 1 a plot of $I_r$ and $I_p$ for the singlet state of the He series.
$I_p$ and $I_r$ are largest for smaller $Z$ in agreement with the argument that electron correlation is largest for smaller values of the nuclear charge. 

One also observes that  $I_r$ and $I_p$ both decrease as Z increases, i.e. there is no complementarity between the two (as in the case of the entropies). $I_p > I_r$ for all the systems studied, thus there is more correlation or information in momentum space as compared to position space. Note also that the difference between the two measures decreases as Z increases. That is, for the more highly correlated systems (Z small), there is a larger difference between the mutual information (correlation) in $r$-space and in $p$-space. In the asymptotic limit as $Z \rightarrow \infty$, the wave function can be written in product form, and both $I_r$ and $I_p$ go to zero.

In the inset of Fig. 1, we plot $I_r$ and $I_p$ for the triplet(1s2s) state of the He series. First, on comparison to the singlet state plot, we notice that both the $I_r$ and $I_p$ values are larger in the triplet case, than in the singlet. This is consistent with the fact that there is more correlation (Fermi plus Coulomb) in the excited triplet state than in the singlet ground state (Coulomb). In the asymptotic limit, the values of $I_r$ and $I_p$ are those corresponding to a non-interacting system which we shall introduce in the following.

The difference between the singlet and triplet states is that in the triplet, $I_p < I_r$, for the more highly correlated systems. At $Z = 5$, the relative order changes, and $I_p > I_r$ as in the singlet case. This is significant since it suggests that the relative ordering of spatial and momentum correlations, or the extent of pairing in each space, depends on the strength of the correlation.  The question is now, Are these differences related to Coulomb correlation, Fermi correlation, or the influence of one of them on the other ? 

A first step would be to eliminate the contribution from Fermi correlation in the triplet state by introducing a reference system, similar to the Hartree-Fock reference that is used in the definition of the correlation energy in quantum chemistry. One can do this by considering non-interacting (NI) triplet states in Eq. (\ref{hewfnst}), where the variational parameters are set 
equal to $Z_1^{\prime}=Z$ and $Z_2^{\prime}=Z/2$ . One could thus define $I_r^{\prime}$ and $I_p^{\prime}$ as
\begin{equation}
I_r^{\prime}= I_r - I_r^{NI} \hspace{2 cm} I_p^{\prime}= I_p -I_p^{NI}.
\label{iprime}
\end{equation}
%
$I_r^{\prime}$ and $I_p^{\prime}$ values should then contain no Fermi correlation, this being subtracted out in the reference system. For these non-interacting systems, $I_r^{NI}$ and $I_p^{NI}$ are constant throughout the series with values of 0.50 and 0.51 respectively. Thus 
$I_p^{NI}$ $>$ $I_r^{NI}$, which is reflected in the use of the term Fermi pressure to describe the Fermi correlation in these states which correspond to the asymptotic limit of large Z.

$I_r^{\prime}$ and $I_p^{\prime}$ are plotted in Fig. 2. With the introduction of the non-interacting reference, the difference between the two measures decreases with Z, as in the case of the singlet systems, and more highly correlated systems are now characterized by a larger difference between the two measures.
However, even with the use of the reference, $I_p^{\prime} < I_r^{\prime}$, opposite to the singlet case. This result suggests a rather complex interplay between Fermi and Coulomb correlation. Another interpretation of these results is that Fermi and Coulomb correlation are both characterized by $I_p$ $>$ $I_r$, in these systems. What is interesting are situations when the two are present and the correlation is strong, which then provokes $I_r$ $>$ $I_p$.

\subsection{Information entropies}

We begin the discussion in this section by examining the relative magnitudes of $S_{\rho}$ and $S_{\pi}$ in hydrogenic atoms, that is, the delocalization due to the effect of the nuclear potential. Simple formulas exist for these entropies \cite{guevarapra} and their behavior is presented as an inset in Fig. 3. One notices that $S_{\rho}$ $>$ $S_{\pi}$ for the hydrogen atom, while for Z $>$ 1, $S_{\rho}$ $<$ $S_{\pi}$. Thus there is a crossover point where the charge density is the most delocalized of the series. We determined that the actual value of Z at the crossover point, where $S_{\rho}$ $=$ $S_{\pi}$, is $\approx 1.33$. The interpretation of this point is that the localization (or delocalization), from the entropic standpoint, is equal in both spaces.

Entropy densities \cite{guevarajcp2,chatz,atre} have been recently introduced into the literature. These are defined in position space as
\begin{equation}
S_{\rho}(r)= -4\pi r^2 \rho (r) \ln \rho (r),
\label{locshanr}
\end{equation}
and in momentum space
\begin{equation}
S_{\pi}(p)= -4\pi p^2 \pi (p) \ln \pi (p).
\label{locshanp}
\end{equation}
Regions of negative local entropies in atomic systems have been associated with localization while positive regions correspond to delocalization \cite{guevarajcp2}. We plot  position space entropic densities of hydrogenic atoms in Fig. 3 since they offer the best perspective to examine the changes. One notes that the crossover point (Z=1) is characterized by delocalization throughout the $r$-space. As Z increases, and one passes this crossover point, there is a region of localization close to the nucleus.

We next turn our attention to plots of $S_{\rho}$, $S_{\pi}$ and $S_{\Gamma}$, $S_{\Pi}$ for the helium triplet systems in Figs. 4 and 5. The singlet systems plot of $S_{\Gamma}$, $S_{\Pi}$ have been presented \cite{guevarajcp1} with the result that $S_{\rho}$ and $S_{\Gamma}$ decrease while $S_{\pi}$ and $S_{\Pi}$ increase, with increasing Z. This behavior is presented as insets in Figs. 4 and 5. One difference is that the $S_{\rho}$ and $S_{\Gamma}$ values are larger in the triplet than in the singlet, i.e. a larger correlation in the triplet induces a greater delocalization of the charge density, relative to the singlet case. The result of this delocalization is that $S_{\pi}$ and $S_{\Pi}$ values are smaller in the triplet, due to the relative localization of the momentum density, as a consequence of the uncertainty principle.

The most striking difference between the singlet and the triplet case (comparing Figs. 4 and 5 to their insets), is that there is a transition or crossover point in the triplet, at small Z (relatively large correlation), which is not present in the singlet systems that were studied. For the singlet, $S_{\pi} > S_{\rho}$, $S_{\Pi} > S_{\Gamma}$, for all Z, however for the triplet(Z=2), $S_{\pi} < S_{\rho}$, $S_{\Pi} < S_{\Gamma}$, which then reverts to the singlet behavior for larger Z. We note that the singlet state H$^-$ ion at the Hartree-Fock(HF) level \cite{clementi} yields $S_{\pi} < S_{\rho}$, $S_{\Pi} < S_{\Gamma}$. Thus there is also a crossover in the singlet state systems at the HF level, but at different Z. The function in Eq. (\ref{hewfns}) does not yield a bound state solution for H$^-$. It is also relevant that the crossover point in the one- and two-electron entropies of non-interacting singlet state systems would be the same as that observed for the one-electron entropies in hydrogenic atoms.
The main point here is that delocalization in position space is the fundamental phenomenon, whether it is directly due to the nuclear potential, or through electron correlation. 

It is plausible to expect that coulomb correlation effects change the actual Z value at the crossover point. For example, the exact value of the crossover point in the two-electron entropies of the non-interacting triplet state series, is $Z\approx 2.50$, while the value for the interacting systems obtained from interpolating the data in Fig. 5 is $Z\approx 2.85$. The result that the Z-value in the interacting case is larger than the non-interacting one is not surprising if one takes into the account the effect of shielding, i.e. electron correlation provides additional impetus for the delocalization in $r$-space.  

From these observations, one may conclude that there is a fundamental re-organization of the densities at both the one- and two-electron levels, due to the uncertainty principle. For the more highly correlated (Z=2) triplet system, there is a relative localization of the momentum density with respect to the charge density while for the lesser correlated systems (Z $>$ 2), there is a relative broadening of the momentum density with respect to the charge density. 

We further investigate this behavior by plotting 
in Figs. 6 and 7 entropy densities  in position and  momentum space
for Z=2-4, that is, around  the transition or crossover point in the triplet state systems. In $r$-space, the highly correlated member (Z=2) is characterized by delocalization, and localization at small $r$ is introduced as Z increases. There is also  structure at larger $r$ in these systems.  

Most significant is the triplet $p$-space entropy density plot in Fig. 7, where the highly correlated member is characterized by a strong localization, at small $p$. As one moves to higher Z (switching off the correlation), this localization disappears. The differences among the entropy profiles is further evidence of the re-organization of the momentum density at all values of $p$.

The relationship between localization and correlation has recently been discussed in terms of information densities \cite{sagar}. Thus one could inquire as to how correlation (Fermi and Coulomb) influences the localization, or as to what is the source of the localization in $p$-space.
A $p$-space information density, or local information, which may be interpreted as a local correlation measure in momentum space,  can be defined analogous to the position space one \cite{sagar} as
\begin{equation}
I_p(p)= {\ln[{N \over N-1}] \over N}I(p) + {S_{\pi}(p) \over N} + {16\pi^2p^2 \over N(N-1)}
[\int p_1^2\Pi (p,p_1)\ln \Pi(p,p_1)dp_1 - \int p_1^2 \Pi(p,p_1) \ln \pi(p_1) dp_1],
\label{ip}
\end{equation}
where the radial momentum distribution is 
\begin{equation}
I(p)=4\pi p^2 \pi (p).
\label{radmom}
\end{equation}

We plot in Fig. 8, $I_p(p)$ for the first three members of the triplet He series. One finds that the crossover point (Z=2) is characterized by stronger correlation (higher peak) in the small $p$ region, which diminishes as the correlation is switched off (larger Z). This strong correlation occurs in the same region as the localization. Small $p$ corresponds to large $r$ due to the inverse nature of the Fourier transform relationship. There is also a second peak in $I_p(p)$ at larger $p$, smaller in comparison to the first, but higher in comparison to the lesser correlated systems.
Hence our conclusion from these results is that the crossover or transition point in these systems is related to correlation in momentum space, resulting from a localized momentum density, and due to the uncertainty principle, since the position space charge density is delocalized as a result of electron correlation.

At selected points in the paper, we have commented on the 
similarities between our work and that in quantum information 
theory. However, there are differences that need to be taken into 
account. The density matrices used in quantum information theory are 
usually expressed in a finite basis set 
(frequently pertaining to models of spin systems) while the 
formulation of the atomic problem involves continuous variables 
since the density matrices and densities are represented in terms of 
the eigenfunctions of the position and momentum operators, 
respectively. Quantum information theory as related to continuous 
variables has been applied to Gaussian states in quantum optics 
where the formulation of the problem is in terms of the joint 
position-momentum Wigner function. Our work has studied electron 
correlation utilizing information theoretic concepts, and the 
densities in position and momentum space (the marginals of the 
Wigner function). That is, we have studied the impact of electron 
correlation by examining the correlation between the variables $r_1$ 
and $r_2$, and $p_1$ and $p_2$, from $I_r$ and $I_p$. The impact of 
electron correlation on entanglement 
(quantum correlations) and non-separability, resides in the correlation 
between the 
position and momentum space variables provided by the uncertainty 
relation in Eq.(\ref{unceri}), and the relationship between $I_r$ 
and $I_p$. 
One can re-write Eq. (\ref{unceri}) in terms of conditional
entropies \cite{sagar} as,

\begin{equation}
I_r + I_p = S^u_{\rho} + S^u_{\pi} - [S({\bf{r}}_1|{\bf{r}}_2)+ S({\bf{p}}_1|{\bf{p}}_2)] \geq 0.
\label{unceri1}
\end{equation}
Thus $I_r + I_p$ is greater than zero, when the sum of the conditional entropies is smaller than the standard entropy sum. This may
be interpreted as evidence of a quantum correlation since measurement of one of the particles induces that the entropy sum,
as measured by the sum of the conditional entropies, is lesser (more correlated) than the standard entropy sum.
Another avenue lies in studying the problem in phase- 
space with two-electron phase-space distributions. The relationship 
between the entropic and information 
uncertainty relations and phase-space distributions has been 
commented on \cite{sagar,guevarajcp1}.


%

\newpage
%
%

%
%

\begin{figure}[tb]
\begin{center}
   \includegraphics*[width=4.in,angle=-90]{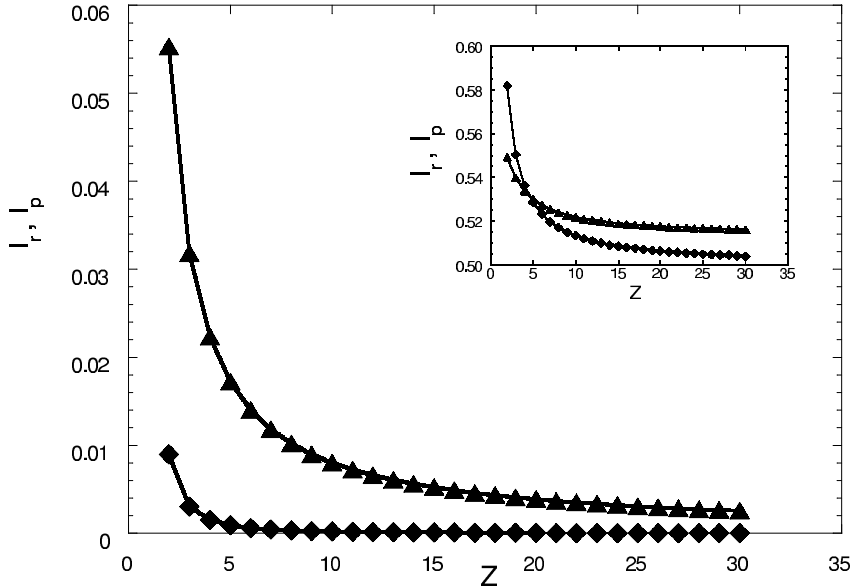}
    \caption{\label{fig1} Plot of $I_r$(diamonds) and $I_p$(triangles)
      for the singlet state members of the He isoelectronic series ($2
 \leq Z \leq 30$). The inset is a plot of $I_r$(diamonds) and $I_p$(triangles)  for the triplet state members.}
\end{center}
\end{figure}


\begin{figure}[tb]
\begin{center}
   \includegraphics*[width=4.in,angle=-90]{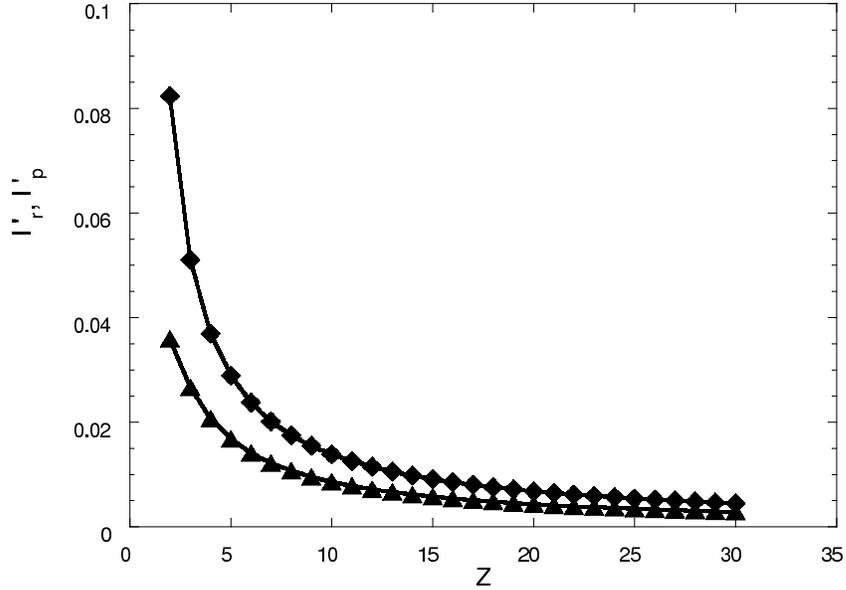}
    \caption{\label{fig2} Plot of $I_r^{\prime}$(diamonds) and $I_p^{\prime}$(triangles) for 
the triplet state members of the He isoelectronic series ($2 \leq Z \leq 30$).}
\end{center}
\end{figure}


\begin{figure}[tb]
\begin{center}
   \includegraphics*[width=4.in,angle=-90]{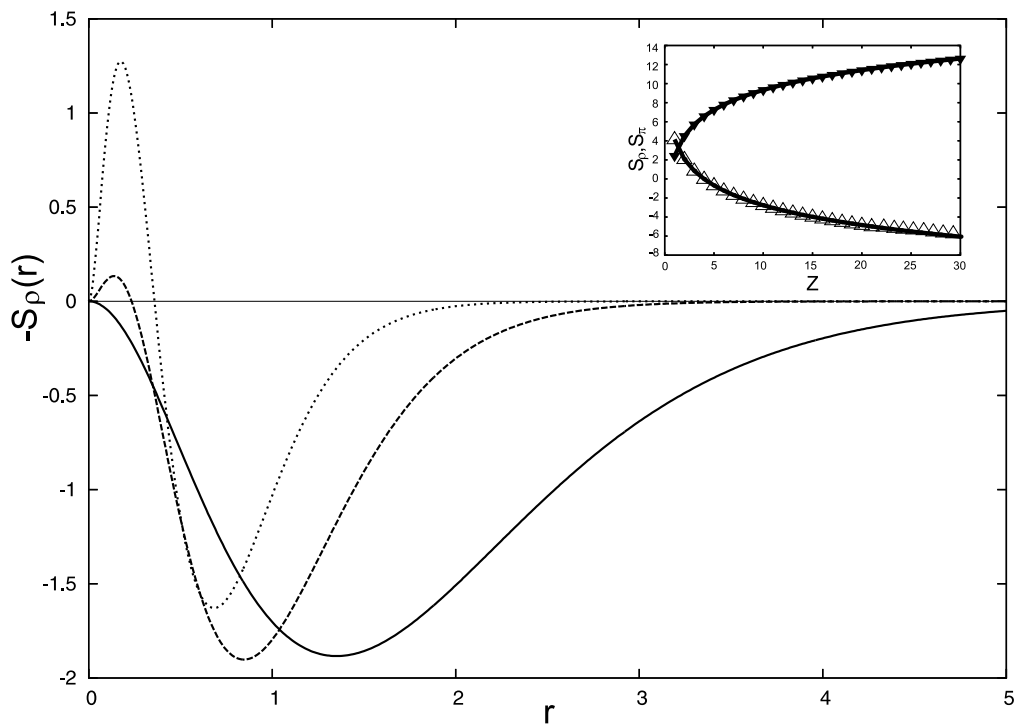}
    \caption{\label{fig3}Plot of $-S_{\rho}(r)$ for the Z=1(solid), 2(dash), 
3(dot) hydrogenic atoms. The inset is a plot of $S_{\rho}$(unfilled triangles)
and $S_{\pi}$(filled triangles) for $1 \leq Z \leq 30$. }
\end{center}
\end{figure}


\begin{figure}[tb]
\begin{center}
   \includegraphics*[width=4.in,angle=-90]{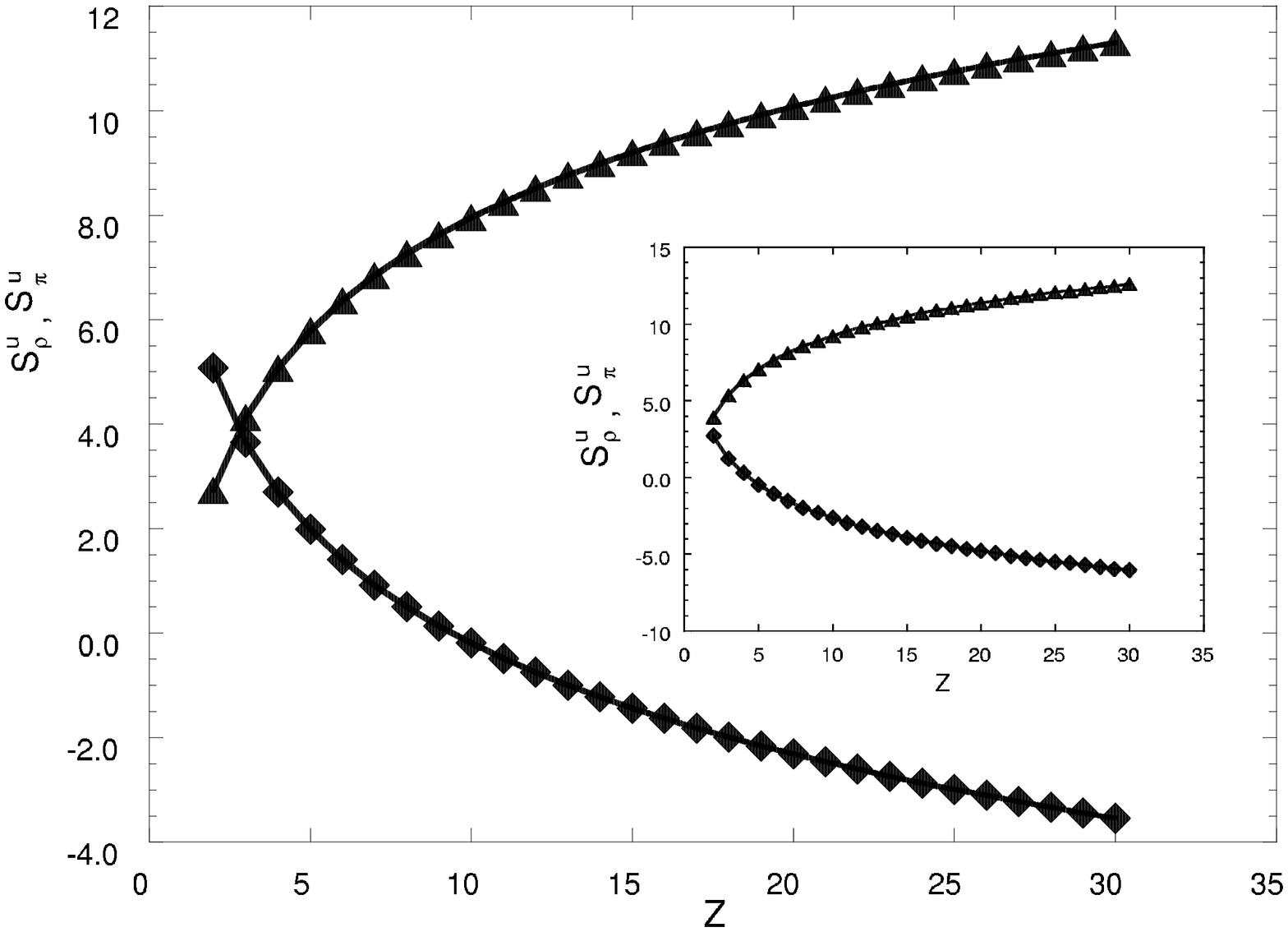}
    \caption{\label{fig4}Plot of $S^u_{\rho}$(diamonds) and $S^u_{\pi}$(triangles) for the 
triplet state
members of the He isoelectronic series ($2 \leq Z \leq 30$). The inset contains 
$S^u_{\rho}$(diamonds) and $S^u_{\pi}$(triangles) of the singlet state 
members. }
\end{center}
\end{figure}


\begin{figure}[tb]
\begin{center}
   \includegraphics*[width=4.in,angle=-90]{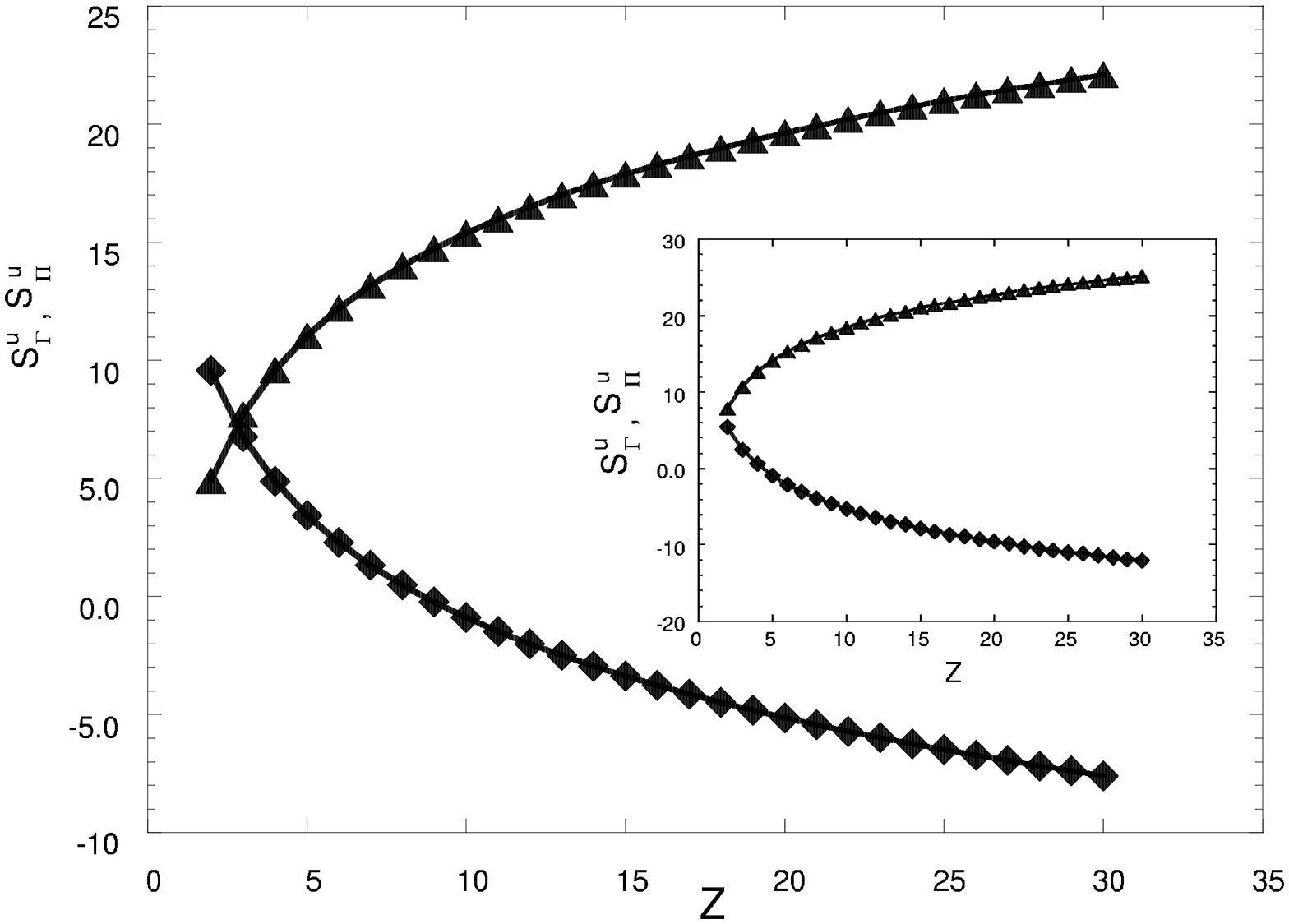}
    \caption{\label{fig5}Plot of $S^u_{\Gamma}$(diamonds) and $S^u_{\Pi}$(triangles)
for the triplet state members of the He isoelectronic series. The inset contains 
$S^u_{\Gamma}$(diamonds) and $S^u_{\Pi}$(triangles)  of the singlet state 
members. }
\end{center}
\end{figure}


\begin{figure}[tb]
\begin{center}
   \includegraphics*[width=4.in,angle=-90]{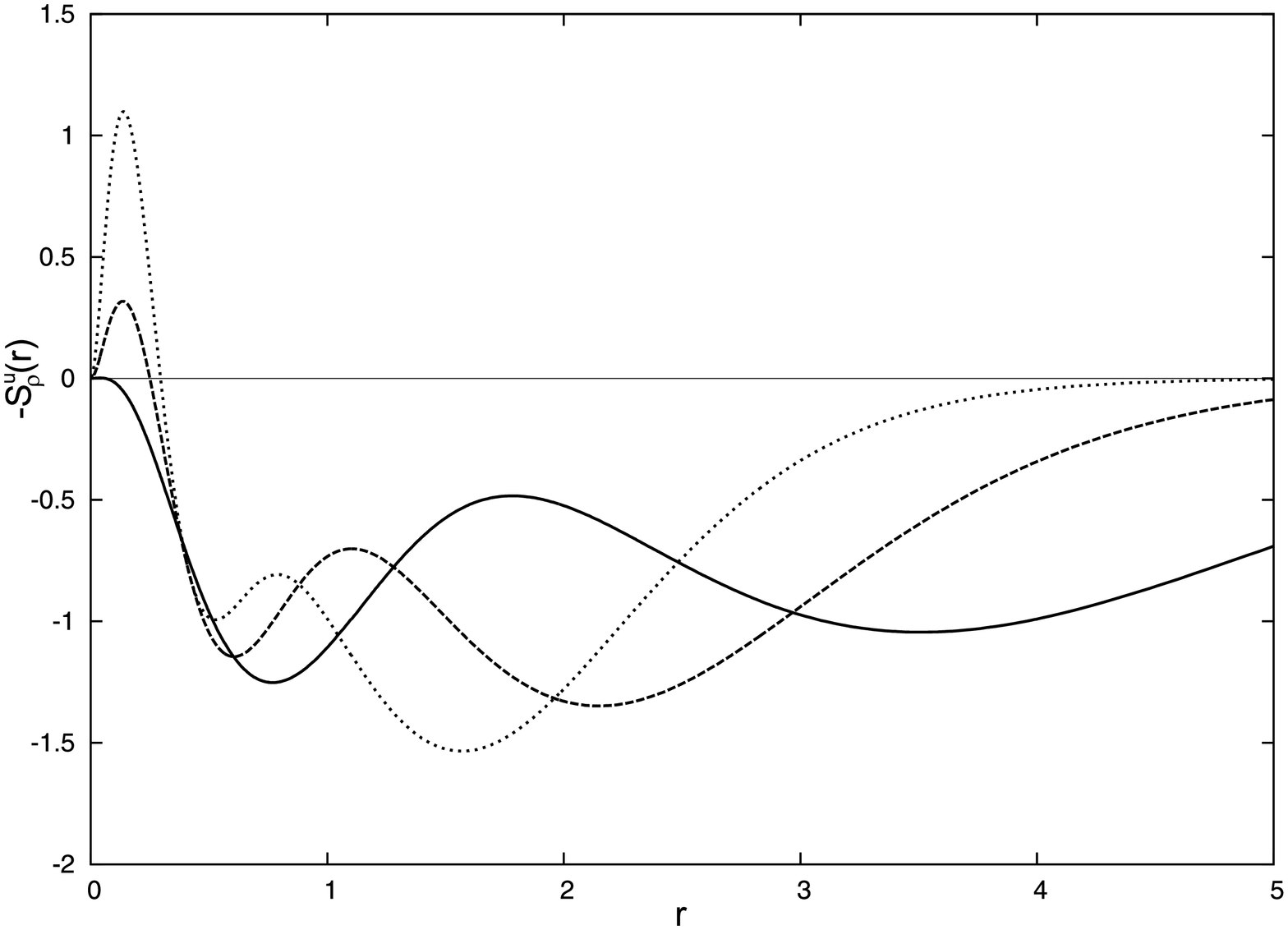}
    \caption{\label{fig6}Plot of $-S^u_{\rho}(r)$ for the Z=2(solid),
      3(dash), 4(dot) members of the triplet
 state He series.}
\end{center}
\end{figure}


\begin{figure}[tb]
\begin{center}
   \includegraphics*[width=4.in,angle=-90]{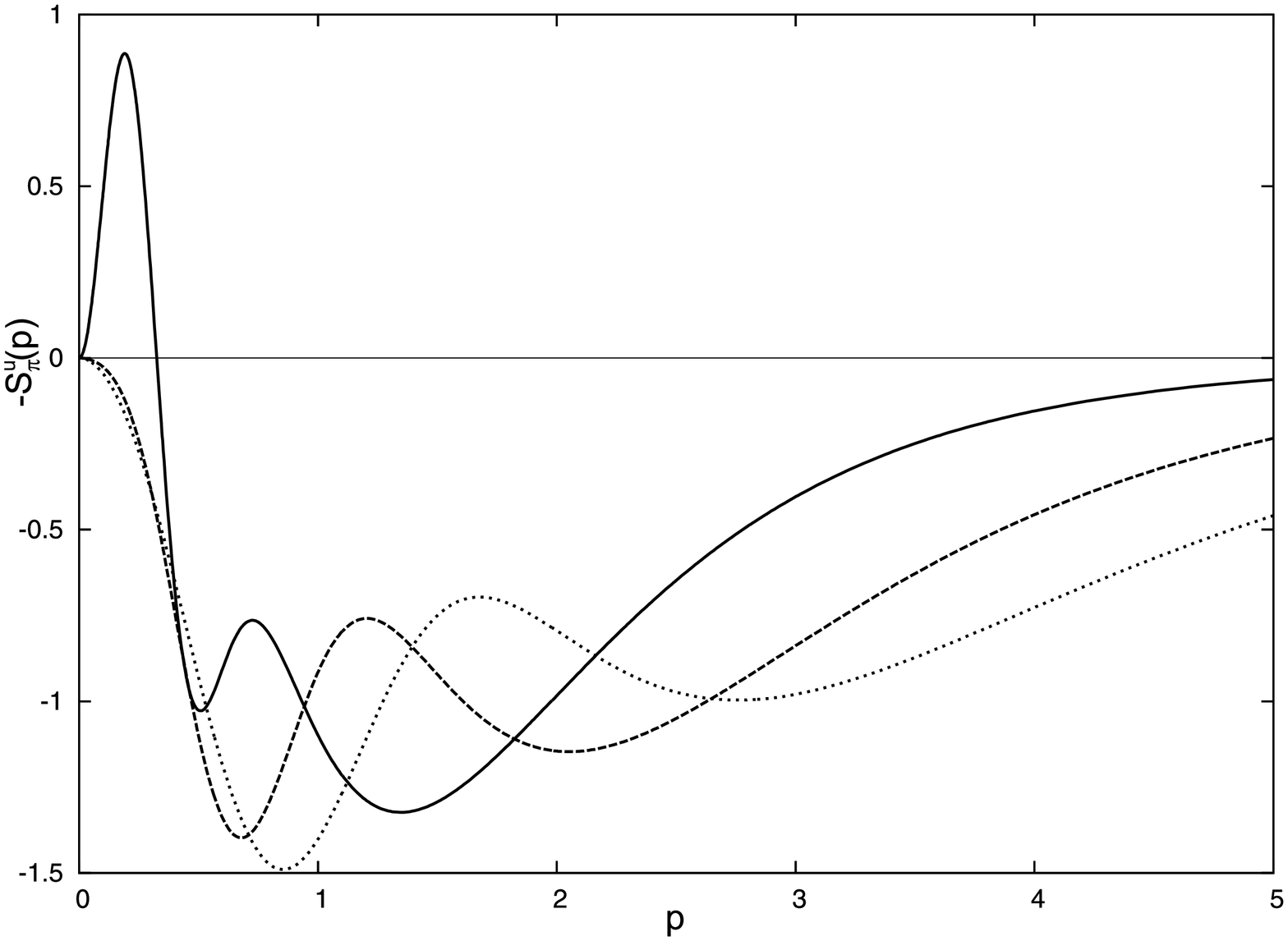}
    \caption{\label{fig7}Plot of $-S^u_{\pi}(p)$ for the Z=2(solid),
      3(dash), 4(dot) 
members of the triplet state He series.}
\end{center}
\end{figure}


\begin{figure}[tb]
\begin{center}
   \includegraphics*[width=4.in,angle=-90]{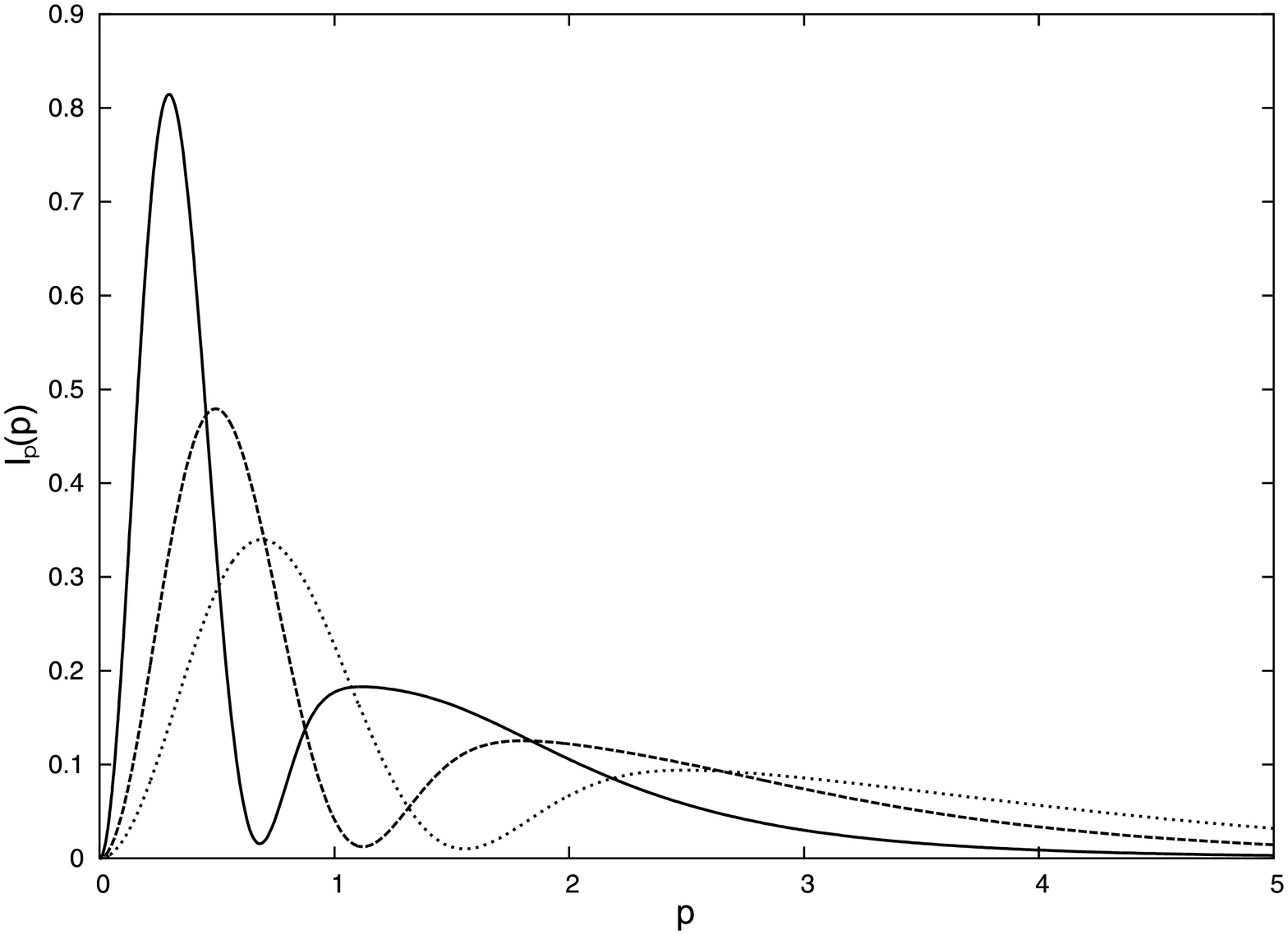}
    \caption{\label{fig8}Plot of $I_p (p)$ for the Z=2(solid), 3(dash), 4(dot) members of the triplet state He series.}
\end{center}
\end{figure}



\section{Conclusions}

The concepts of local and non-local information 
are discussed in terms of mutual 
information and entropic measures in position and 
momentum space. We ask the question 
of whether  the relative magnitude of the spatial 
correlations, as compared to the 
momentum correlations, depends on the strength of 
the electronic correlations. 
We show this to be affirmative for the triplet 
state helium isoelectronic series 
by examining the position and momentum space 
mutual information measures.  
Furthermore, 
the nuclear potential and electron correlation are responsible for 
a transition or crossover point at small Z, where the one and two-electron Shannon entropies in the momentum space  are smaller than the position space ones. This localization of the momentum density is due to the uncertainty principle. 
When Z increases and correlation is switched off, it is the momentum space Shannon entropies which are now larger than the position space ones. An analysis of the entropic densities in momentum space reveals that the crossover point is characterized by a strongly localized momentum density at small values of $p$. As correlation is switched off, the localization disappears. The information density in momentum space demonstrates that this localization is associated with strong correlation, at small $p$.

%


\section{ACKNOWLEDGMENTS}
The authors thank  the Consejo
Nacional de Ciencias y Tecnologia (CONACyt) and the PROMEP program of
the Secretario de Educaci\'{o}n P\'{u}blica in M\'{e}xico for support. 

\newpage

%


\end{document}